\documentclass{article}

\usepackage{microtype}
\usepackage{graphicx}
\usepackage{subfigure}
\usepackage{booktabs} 
\usepackage{fancyhdr}
\usepackage{amsmath}
\usepackage{amsfonts}
\usepackage{hyperref}
\usepackage{listings}
\usepackage{xcolor}

\usepackage[accepted]{mlsys2025}

\newcommand{\final}{0} 
\newcommand{\karthik}   [1]{{{\color{red}(karthik) #1}}}
\ifthenelse{\equal{\final}{1}}{\renewcommand{\karthik}[1]{}}{}
\newcommand{\mansi}   [1]{{{\color{purple}(mansi) #1}}}
\ifthenelse{\equal{\final}{1}}{\renewcommand{\mansi}[1]{}}{}
\newcommand{\muhammad}   [1]{{{\color{orange}(muhammad) #1}}}
\ifthenelse{\equal{\final}{1}}{\renewcommand{\muhammad}[1]{}}{}
\newcommand{\sonali}   [1]{{{\color{green}(sonali) #1}}}
\ifthenelse{\equal{\final}{1}}{\renewcommand{\sonali}[1]{}}{}
\newcommand{\ganesh}   [1]{{{\color{blue}(ganesh) #1}}}
\ifthenelse{\equal{\final}{1}}{\renewcommand{\ganesh}[1]{}}{}

\begin{document}

\twocolumn[
\mlsystitle{Optimizing Attention on GPUs by Exploiting GPU Architectural NUMA Effects}




\begin{mlsysauthorlist}
\mlsysauthor{Mansi Choudhary\textsuperscript{1 *}}{}
\mlsysauthor{Karthik Sangaiah\textsuperscript{2}}{}
\mlsysauthor{Sonali Singh\textsuperscript{2}}{}
\mlsysauthor{Muhammad Osama\textsuperscript{2}}{}
\mlsysauthor{Lisa Wu Wills\textsuperscript{3}}{}
\mlsysauthor{Ganesh Dasika\textsuperscript{2}}{}
\end{mlsysauthorlist}

\mlsyscorrespondingauthor{Mansi Choudhary}{mc846@duke.edu}
\mlsyscorrespondingauthor{Karthik Sangaiah}{karthik.sangaiah@amd.com}

\vskip 0.15in
{\centering\small
\textsuperscript{1}Department of ECE, Duke University, Durham, USA\\
\textsuperscript{2}Advanced Micro Devices Inc., Santa Clara, USA\\
\textsuperscript{3}Department of Computer Science, Duke University, Durham, USA\\
}


\vskip 0.3in
\begin{abstract}

The rise of disaggregated AI GPUs has exposed a critical bottleneck in large-scale attention workloads: non-uniform memory access~(NUMA). As multi-chiplet designs become the norm for scaling compute capabilities, memory latency and bandwidth vary sharply across compute regions, undermining the performance of traditional GPU kernel scheduling strategies that assume uniform memory access. We identify how these NUMA effects distort locality in multi-head attention~(MHA) and present Swizzled Head-first Mapping, a spatially-aware scheduling strategy that aligns attention heads with GPU NUMA domains to exploit intra-chiplet cache reuse. On AMD’s MI300X architecture, our method achieves up to 50\% higher performance over state-of-the-art attention algorithms using conventional scheduling techniques and sustains consistently high L2 cache hit rates of 80-97\%. These results demonstrate that NUMA-aware scheduling is now fundamental to achieving full efficiency on next-generation disaggregated GPUs, offering a path forward for scalable AI training and inference.
\end{abstract}
]

\let\thefootnote\relax 
\footnotetext{\textsuperscript{*}Work completed during an internship at Advanced Micro Devices Inc. Correspondence to: Mansi Choudhary \textless{}mc846@duke.edu\textgreater{}, Karthik Sangaiah \textless{}karthik.sangaiah@amd.com\textgreater{}.}




\section{Introduction}

\begin{figure*}
    \centering
    \includegraphics[width=0.9\linewidth]{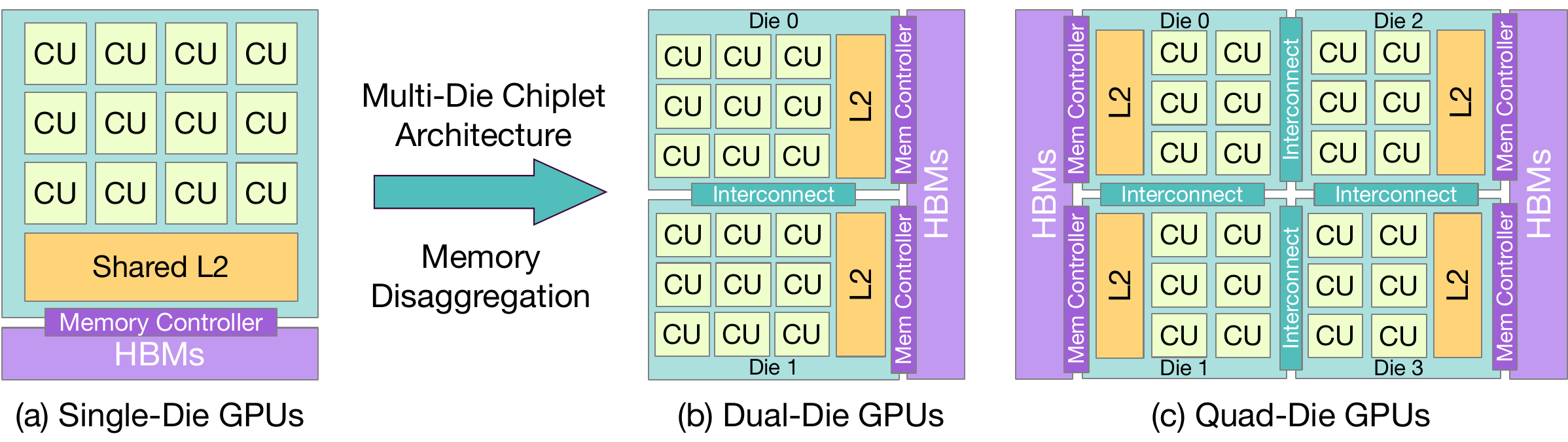}
    \caption{Evolution of GPU architectures toward disaggregated memory hierarchies. (a) Traditional single-die GPU (e.g., NVIDIA A100~\cite{nvidia_a100}, H100~\cite{nvidia_h100}; AMD MI200 series~\cite{amd_mi200}) with unified L2 cache shared across all compute units (CUs), providing uniform memory access. (b) Dual-die chiplet architecture (e.g., NVIDIA Blackwell~\cite{nvidia_blackwell}, Rubin~\cite{nvidia_rubin}) with interconnects between dies. (c) Quad-die chiplet architecture (e.g., NVIDIA Rubin Ultra~\cite{nvidia_rubin_ultra}, AMD MI300~\cite{amd_mi300} series) with further disaggregation. Each die has dedicated compute units, L2 cache, and memory controllers connected to HBM stacks. While AMD and NVIDIA both employ multi-die designs, the degree to which NUMA effects are exposed to software varies by implementation. NVIDIA's Blackwell maintains full cache coherency between the dies, abstracting the NUMA effects at the hardware level, whereas AMD's MI300X explicitly exposes NUMA characteristics, enabling architecture-aware optimizations.}
    \label{fig:chiplet}
\end{figure*}

The growth of machine learning and generative AI has created unprecedented demand for highly efficient compute systems. As models scale to billions of parameters, train on increasingly large datasets, and handle massive inference workloads, the computational requirements have grown exponentially, driving the need for more powerful and efficient AI accelerators. This relentless scaling pressure has fundamentally reshaped how ML accelerator architectures, including GPUs, are designed and manufactured.

To meet these scaling demands while managing manufacturing costs and yields, AI accelerator architectures, including AI GPUs, have evolved toward increasingly disaggregated designs. Modern AI GPUs feature distributed memory hierarchies with multiple memory controllers and cache structures spread across large die areas or multiple chiplets. This disaggregation, whether through larger monolithic dies with distributed resources or chiplet-based multi-die integration, fundamentally alters the organization of the memory subsystem. Unlike earlier GPU generations with relatively uniform memory access patterns, these scaled architectures introduce non-uniform memory access~(NUMA) effects, where memory access latencies and bandwidth characteristics vary significantly based on the spatial relationship between compute units and memory resources. Chiplet-based designs represent a natural evolution in this trajectory, offering improved manufacturing yields and enhanced scalability while making NUMA effects particularly pronounced due to inter-die communication overhead.

The Transformer architecture~\cite{attn} has enabled remarkable advances in large language models~(LLMs) and generative AI. The Attention mechanism, the fundamental component of Transformers, has become a primary bottleneck in generative AI workloads due to its quadratic time and memory complexity as sequence lengths grow. This bottleneck has fueled extensive optimization efforts over recent years~\cite{flashattention, flashattention2, flashattention3, leanattention}, with approaches ranging from algorithmic improvements~\cite{flashattention, flashattention2, leanattention} to hardware-specific optimizations~\cite{flashattention3}.

However, existing attention optimizations largely assume uniform memory access patterns and do not account for the NUMA effects present in modern scaled GPU architectures. Rather than viewing NUMA characteristics as obstacles to overcome, we recognize them as creating new opportunities for performance optimization through spatially-aware kernel design. To demonstrate this approach, we focus our study on AMD's MI300X architecture, which represents one of the most advanced implementations of chiplet-based design in current-generation AI accelerators. The MI300X features eight XCDs (Accelerator Complex Dies), each with dedicated compute units, L2 cache, and memory controllers connected to independent HBM stacks. This multi-chiplet configuration with distributed memory subsystems makes NUMA effects particularly pronounced, providing an ideal testbed for developing and validating spatially-aware optimization techniques. As disaggregated and chiplet-based architectures become increasingly prevalent across the industry for scaling AI workloads, insights gained from optimizing for such designs will have broad applicability to future generations of GPUs and other AI accelerators.

This focus on NUMA-aware optimization builds upon prior work demonstrating significant benefits from architecture-aware kernel design. For instance, spatially-aware mapping strategies for GEMM operations on the MI300X have shown dramatic improvements in cache utilization, with L2 cache hit rates increasing from 43\% to 92\% when accounting for NUMA characteristics~\cite{tensile}. These results underscore the critical importance of architecture-aware optimization and motivate our investigation of similar techniques for attention kernels, which represent another critical bottleneck in modern AI workloads.

In this paper, we introduce \textbf{Swizzled Head-first Mapping}, a novel, spatially-aware technique that exploits the inherent spatial locality patterns in Attention computations to mitigate NUMA effects and maximize cache utilization in chiplet-based GPU architectures. Our approach builds upon the success of spatially-aware GEMM optimizations by identifying and exploiting analogous locality patterns within the Attention mechanism.


In this work, we present the following contributions:
\begin{itemize}
    \item Analysis of NUMA effects in chiplet-based multi-die GPU architectures and their implications for kernel performance
    \item Characterization of spatial locality patterns inherent in attention mechanisms (forward and backward variants) and identification of opportunities for memory hierarchy optimization
    \item Design and implementation of \textbf{Swizzled Head-first Mapping}, a novel spatially-aware attention optimization technique that aligns computational work with NUMA boundaries to maximize cache efficiency
    \item Comprehensive performance evaluation comparing spatially-unaware approaches to our proposed mapping, demonstrating up to 50\% higher performance over state-of-the-art attention algorithms using conventional scheduling techniques and consistently sustaining high L2 cache hit rates of 80-97\%.
\end{itemize}

\section{Background}

\subsection{NUMA Effects in Modern GPU Architectures}
\label{sec:NUMA}
Modern GPU architectures designed for AI workloads have increasingly adopted disaggregated memory hierarchies to achieve the necessary scale in compute capacity and memory bandwidth. Whether through large monolithic designs with distributed resources or chiplet-based multi-die integration, these architectures depart from traditional unified cache designs in which all compute units share common cache structures. This disaggregation introduces NUMA effects, where memory access latencies and bandwidth characteristics vary based on the spatial relationship between compute units and the memory resources they access.

NUMA effects manifest as a hierarchy of memory access costs within the GPU. Local accesses, where compute units access memory resources physically proximate to them, benefit from high-bandwidth, low-latency paths through nearby private caches. In contrast, remote accesses that must traverse longer distances or cross interconnect boundaries experience substantially higher latencies and reduced effective bandwidth. Furthermore, distributed cache hierarchies mean that data cached in one region provides no benefit to compute units in other regions, effectively fragmenting the total cache capacity. Workloads that fail to account for these spatial characteristics suffer from reduced cache hit rates and poor memory bandwidth utilization.

This transition from unified to distributed memory hierarchies fundamentally changes optimization priorities for GPU kernels. In traditional architectures with uniform access patterns, temporal locality was the primary optimization target, with the focus on \textit{when} work is scheduled to maximize cache reuse over time. Architectures with NUMA effects introduce spatial dimensions to this optimization space, where both \textit{when} and \textit{where} work is scheduled become equally critical. Performance becomes highly sensitive to the spatial distribution of computational work and data placement, requiring kernel designers to explicitly consider the physical topology of the underlying hardware to achieve optimal performance.

\subsection{Current Mitigations for Spatial NUMA Effects}

\begin{figure}[h]
  \centering
  \includegraphics[width=\columnwidth]{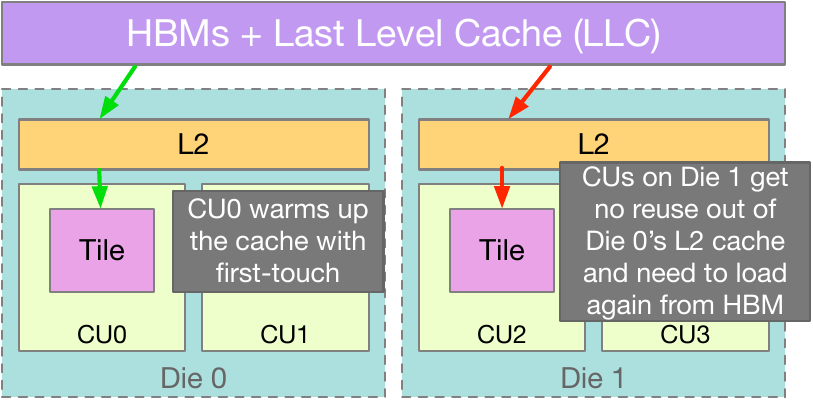}
  \caption{Impact of workgroup scheduling on cache reuse in the multi-die chiplet architecture of AMD MI300X\textsuperscript{TM}. When workgroups processing related tiles that share input data are scheduled to different dies (left: CU0 on Die 0, right: CU2 on Die 1), they cannot benefit from each other's cached data. This cross-die scheduling forces redundant memory fetches from HBM through the shared last-level cache (LLC), as L2 caches are private to each Accelerator Complex Die or XCD. The NUMA effects inherent to this disaggregated architecture make spatially aware workgroup placement critical for maximizing cache efficiency and minimizing memory bandwidth consumption.}
  \label{fig:chunksize1}
\end{figure}

Modern multi-die chiplet GPUs schedule workgroups (WGs) across compute dies using a chunked round-robin policy, where each die (or NUMA domain) receives a small, contiguous batch of WGs before the scheduler advances to the next. On current hardware, this chunk size is set to one (see Figure~\ref{fig:chunksize1}), ensuring load balance and full utilization of aggregate HBM bandwidth. However, this default policy can inadvertently fragment spatial locality when adjacent WGs operate on neighboring regions of the same tensor but are dispatched to different dies.

Because this mapping strategy is implemented in the driver and subject to change across GPU generations, programmers cannot rely on fixed scheduling behavior. Instead, kernels must incorporate mutable, algorithmic mapping logic that adapts at runtime. One widely adopted technique to address this challenge is \textit{swizzling}, which refers to the deliberate reordering of how computational work is mapped to hardware execution units to improve memory locality and cache utilization. As illustrated in Figure~\ref{fig:wg-swizzle}, swizzling remaps workgroup IDs so that spatially adjacent tiles are assigned to the same die, thereby exploiting per-die L2 caches and improving locality.

This remapping approach generalizes the cooperative thread array (CTA) or workgroup swizzling techniques widely used in high-performance GEMM libraries such as Tensile~\cite{tensile}, hipBLASLt~\cite{hipBLASlt}, and Triton~\cite{triton-swizzle2d}. In these libraries, swizzling transforms the default linear workgroup assignment into patterns that better align with memory layouts or hardware topology. For instance, matrices may be tiled into 2D blocks to maximize data reuse within shared caches. When applied to workloads on multi-die architectures, such remapping aligns the logical computation order with the physical NUMA topology, mitigating scheduler-induced inefficiencies while maintaining portability across future hardware schedulers. In this work, we extend these techniques to implement spatially-aware mapping patterns for Attention on Instinct GPUs.



\begin{figure}[h!]
    \centering
    \begin{lstlisting}[language=Python, 
                       basicstyle=\footnotesize\ttfamily,
                       breaklines=true,
                       backgroundcolor=\color{gray!10},
                       numbers=left,
                       numberstyle=\tiny,
                       numbersep=5pt,
                       showstringspaces=false,
                       commentstyle=\color{green!50!black},
                       keywordstyle=\color{blue},
                       stringstyle=\color{red}]
@triton.jit()
def swizzle_chiplet(wgid, grid, NUM_XCD: t1_constexpr):
    # Number of wgids/XCD in the new config
    wgids_per_xcd = grid // NUM_XCD

    # Compute local wgid within the XCD
    xcd = wgid % NUM_XCD
    local_wgid = wgid // NUM_XCD

    # New wgid based on the new grouping
    new _wgid = xcd * wgids_per_xcd + local_wgid
    return new_wgid
    \end{lstlisting}
    \caption{Chiplet-aware workgroup ID remapping in Triton. The function transforms the original workgroup ID by determining the target XCD and calculating the new global workgroup ID to achieve spatial locality within chiplet memory domains.}
    \label{fig:wg-swizzle}
\end{figure}

\subsection{The Attention Mechanism}
Attention~\cite{attn} is the central component in the Transformer architecture that has enabled the remarkable capabilities of modern generative AI models. The attention mechanism allows models to selectively focus on the most relevant parts of an input sequence when generating each output token. This capability enables language models to maintain coherent context across long sequences and produce high-quality, contextually appropriate outputs.

The core attention computation involves three key matrices derived from the input: queries ($Q$), keys ($K$), and values ($V$). Given these input sequences and head dimension $d$, the attention output $O$ is computed through a three-step process:
\begin{equation}
\label{eq:fa}
S = QK^{T}, P = softmax(\frac{S}{\sqrt{d}}), O = PV.
\end{equation}

In Equation \ref{eq:fa}, the first step computes attention scores $S$ by taking the dot product between queries and keys, measuring the relevance of each key to each query. These scores are then normalized using the softmax function to produce attention weights $P$, which represent the relative importance of each input position. Finally, these weights are used to compute a weighted average of the value vectors, producing the final output $O$.

During training, gradients must be computed with respect to all inputs. Given the gradient of the loss with respect to the output $dO$, the backward pass computes gradients through the attention computation in reverse order:
\begin{equation}
\label{eq:ba}
\begin{aligned}
dV &= P^T dO \in \mathbb{R}^{N \times d}, \quad dP = dO V^T \in \mathbb{R}^{N \times N}, \\
dS &= \mathrm{dsoftmax}(dP) \in \mathbb{R}^{N \times N}, \\
dQ &= dS K \in \mathbb{R}^{N \times d}, \quad dK = dS^T Q \in \mathbb{R}^{N \times d}.
\end{aligned}
\end{equation}

\begin{figure}[h]
    \centering
    \includegraphics[width=0.9\columnwidth]{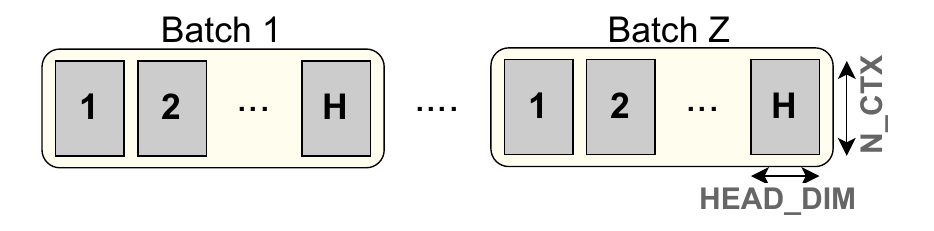}
    \caption{Attention Algorithm Grid for $Q$, $K$, $V$, $O$ tensors. Z$=$batch size, H$=$\# of attention heads, sized by query context length~(N\_CTX) and head dimension~(HEAD\_DIM).}
    \label{fig:attn-grid}
\end{figure}

In practice, Transformer models use multi-head attention (MHA), which splits the model dimension ($d$) across multiple attention heads. Each head operates on a portion of the full dimensionality with separate query, key, and value projections, allowing different heads to capture distinct relationships and attend to various aspects of the input sequence. More recent models have adopted grouped query attention (GQA)~\cite{gqa}, which reduces memory requirements by sharing key and value projections across multiple query heads while maintaining separate query projections. This significantly reduces the memory footprint during inference. Both MHA and GQA create computational grids spanning multiple attention heads and batch items. Each grid cell operates independently with no data reuse between heads or batch items. Figure~\ref{fig:attn-grid} shows this grid, where each head spans a query length~($\mathrm{N\_CTX}$) and head dimension~($\mathrm{HEAD\_DIM}$).

A key limitation of traditional attention is its quadratic time and memory complexity, especially as sequence length grows. This becomes a bottleneck due to the need to store intermediate tensors ($S$, $P$) in global memory.

\subsubsection{Flash Attention}
FlashAttention~(FA)~\cite{flashattention} addresses this limitation by using an online softmax approach, enabling tiled computation where intermediate blocks of $S$ and $P$ reside in local memory rather than global memory. A cross-tile ``fix-up" step ensures numerical correctness while significantly reducing memory overhead. This tiling strategy applies to both the forward pass, which computes attention outputs, and the backward pass, which computes gradients with respect to $Q$, $K$, and $V$ during training.

\begin{figure*}[t]
    \centering
    \includegraphics[width=0.8\linewidth]{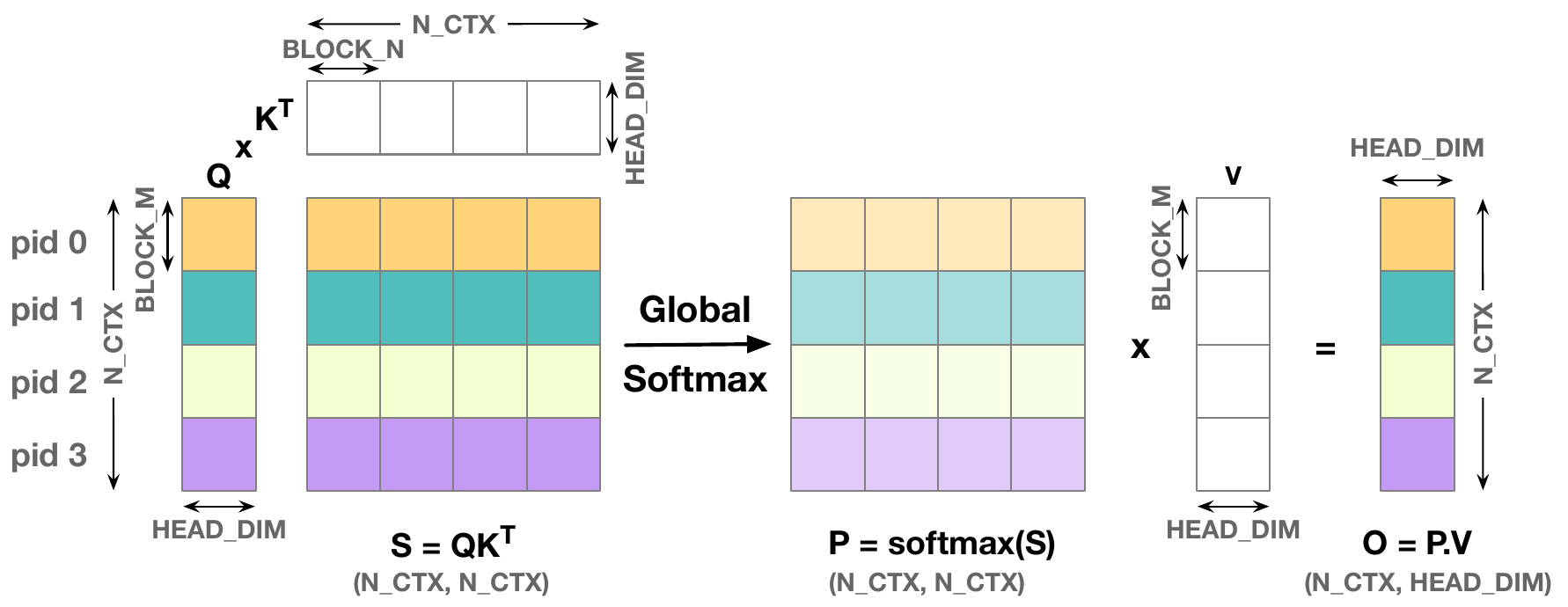}
    \caption{FlashAttention2 Tiled Compute Partitioning Across Workgroups for a Single Attention Head. Query matrix $Q$ is partitioned into row blocks (BLOCK\_M), with each workgroup (pid0-pid3) processing one row block. Each workgroup accesses the complete key matrix $K^T$ and value matrix $V$ to compute attention scores $S = QK^T$, apply softmax to obtain attention weights $P$, and produce output $O = PV$. All workgroups within the same attention head share access to the same $K$ and $V$ tensors, creating natural spatial locality patterns. Matrix dimensions are shown in parentheses, where N\_CTX is the context length and HEAD\_DIM is the head dimension.}
    \label{fig:fa2}
\end{figure*}

FlashAttention2~(FA2)~\cite{flashattention2} further improves performance by parallelizing WGs across the context length for both forward and backward computations. As shown in Figure~\ref{fig:fa2}, $Q$ is partitioned into row blocks processed in parallel, one row block per WG, with each accessing the full $K$ and $V$ tensors. During the backward pass, the gradient computation follows a similar tiling pattern, with each WG computing gradients for its assigned $Q$ block while accessing the complete $K$ and $V$ tensors. This data reuse pattern across both forward and backward passes is central to our optimization for chiplet-based GPUs, as detailed in Section~\ref{sec:mapping}.

\section{FA2 Spatial Locality and Mapping Patterns}
\label{sec:mapping}
This section explores the spatial locality inherent in the FlashAttention2 algorithm and demonstrates how it can be leveraged to optimize performance on chiplet-based GPU architecture, specifically AMD's chiplet-based Instint GPU - MI300X. We first analyze FA2's computational grid structure and examine current methods for mapping this grid to distributed compute dies or XCDs. 
We then introduce a novel optimization, \textbf{Swizzled Head-first Mapping}, that improves data reuse by aligning computational work with the NUMA characteristics of chiplet architectures, thereby enhancing per-XCD cache utilization and overall performance.


\begin{figure}
    \centering
    \includegraphics[width=0.95\linewidth]{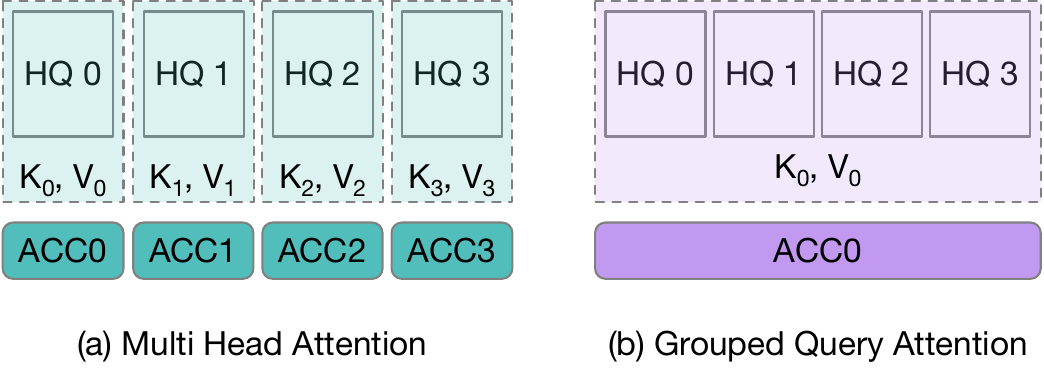}
    \caption{Attention compute cluster (ACC) organization in transformer attention mechanisms. Operations involving the same K and V tensors are co-located within a single ACC for optimal cache utilization. (a) In multi-head attention, each head maintains distinct K, V tensors (e.g., DeepSeek-V3), requiring one ACC per head. (b) In a grouped query attention, multiple heads (four in this figure) share K, V tensors within a group (e.g., Llama, Mistral), requiring one ACC per group.}

    \label{fig:acc}
\end{figure}

\subsection{Spatial Locality in FlashAttention2}
FlashAttention2 exhibits inherent spatial locality patterns in both forward and backward passes, creating opportunities for cache optimization in chiplet-based architectures. As illustrated in Figure~\ref{fig:fa2}, during the forward pass, within a single attention head, each row block of the query matrix $Q$ must access the complete key ($K$) and value ($V$) tensors to compute attention outputs. This access pattern creates natural data-sharing opportunities, as multiple workgroups processing different row blocks of the same attention head (e.g., pid0 to pid3 in Figure~\ref{fig:fa2}) require the same $K$ and $V$ data.

The backward pass follows a similar spatial locality pattern. During gradient computation, each workgroup computing different row blocks of the gradients $dQ$, $dK$, and $dV$ may share the same $Q$, $K$, $V$, and $dO$ tensors within the same attention head. This consistent data sharing pattern across both forward and backward passes amplifies the benefits of cache-aware mapping strategies.

The extent of this data sharing in both the forward and the backward pass depends on the specific attention mechanism employed. In MHA, each attention head maintains its own distinct $K$ and $V$ tensors, meaning workgroups within a single head share data during both forward and backward computation, but workgroups across different heads operate on entirely separate tensors. In contrast, GQA extends this sharing pattern by having multiple query heads share the same key and value tensors, effectively creating larger groups of workgroups that all access the same K and V data during both forward and backward passes.

From a spatial locality perspective, workgroups that share the same input tensors during either forward or backward computation naturally form what we term an \textbf{Attention Compute Cluster (ACC)}. In MHA, each ACC corresponds to all workgroups within a single attention head for both passes. In GQA, each ACC encompasses all workgroups across the grouped heads that share the same key-value tensors during training. The key optimization insight is that co-locating all workgroups within an ACC on the same XCD maximizes L2 cache utilization through two mechanisms: first, shared tensor data of an ACC is loaded into a single XCD's L2 cache and reused across multiple workgroups, preventing cache fragmentation while increasing data reuse; second, this reduces the total number of memory fetches from HBM, as each shared tensor is loaded only once per device rather than redundantly across multiple XCDs.
This approach mirrors successful tiling strategies used in GEMM optimizations, where co-locating workgroups that share input data significantly improves memory bandwidth utilization and compute throughput.

\subsection{Current FA2 Grid-to-XCD Mapping Patterns}
FA2 supports multiple strategies for mapping WGs to XCDs. We categorize existing approaches based on their iteration order and XCD assignment strategy.

\subsubsection{Naive Block-first} The Naive Block-first strategy iterates through all attention heads for each row block before moving to the next block (i.e., completes block0 across all heads, then block1 across all heads, etc.). WGs are assigned to XCDs in round-robin fashion~(e.g., XCD0 gets block0 of HQ0 (Query Head 0), XCD1 gets Block 0 of HQ 1, etc). This splits ACCs across XCDs, reducing cache efficiency. This scheme is a baseline, un-swizzled version of the mapping in section \ref{sec:rbf}.

\begin{figure}[h!]
    \centering
    \includegraphics[width=0.9\linewidth]{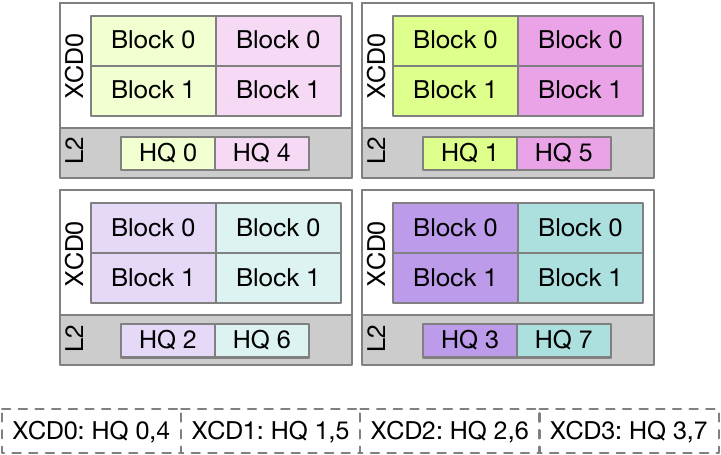}
    \caption{Naive Block-first mapping (illustrated with eight qheads, 128 row blocks, four XCDs): WGs are dispatched block-by-block across all heads in round-robin fashion, splitting ACCs across XCDs. Each unique color corresponds to a unique attention head.}
    \label{fig:naive-block-first}
\end{figure}

\begin{figure}[h!]
    \centering
    \includegraphics[width=0.9\linewidth]{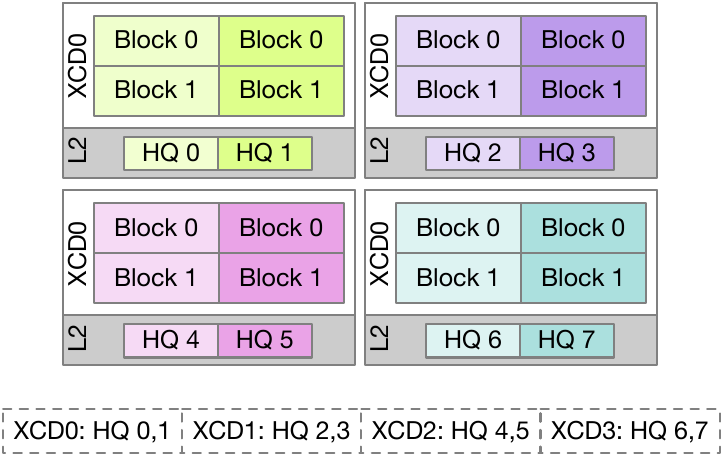}
    \caption{Swizzled Block-first mapping (eight qheads, 128 row blocks, four XCDs): Block-first iteration with GQA-aware swizzling to co-locate grouped heads within XCDs. This scheme works best when GQA groups match XCD count. Each unique color corresponds to a unique attention head.}
    \label{fig:swizzled-block-first}
\end{figure}

\subsubsection{Swizzled Block-first} \label{sec:rbf} The Swizzled Block-first approach retains the block-first iteration order but applies a swizzling technique to co-locate grouped heads in GQA within the same XCD, preserving locality (Figure~\ref{fig:swizzled-block-first}). However, this only maintains locality when the number of GQA groups matches the number of XCDs. For MHA, used in the prefill stage of models like DeepSeek-V3~\cite{deepseekv3}, this strategy causes multiple ACCs per XCD to be served simultaneously, reducing L2 cache efficiency. This scheme is deployed by AMD's AITER~\cite{aiter} repository of high-performance AI operators.

\subsubsection{Naive Head-first} 
\begin{figure}[h!]
    \centering
    \includegraphics[width=0.9\linewidth]{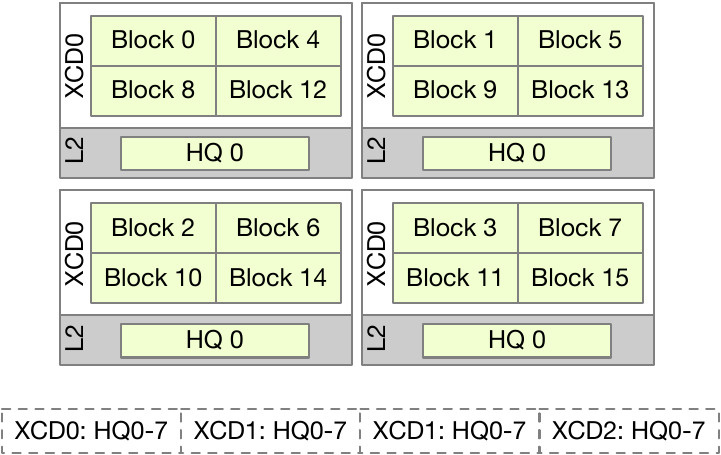}
    \caption{Naive Head-first mapping (eight qheads, 128 row blocks, four XCDs): WGs iterate through all blocks of each head sequentially, but round-robin XCD assignment spreads each head across all XCDs. All blocks correspond to the same attention head in this figure.}
\label{fig:naive-head-first}
\end{figure}

The Naive Head-first strategy iterates through all row blocks of a single attention head before moving to the next head (i.e., completes all blocks of HQ 0, then all blocks of HQ 1, etc.). With round-robin XCD assignment, each head's blocks are striped across all XCDs (Figure~\ref{fig:naive-head-first}). While this maintains head-level coherence during iteration, it still splits individual ACCs across XCDs. This scheme is implemented in Triton's default FlashAttention kernel~\cite{triton}.

For illustration, Figures~\ref{fig:naive-block-first}--\ref{fig:naive-head-first} show these strategies for a simplified example with eight query heads, 128 row blocks per head, and four XCDs.

\subsection{Swizzled Head-first Mapping}
To maximize spatial locality on AMD GPUs with disaggregated L2 caches, we leverage a key insight: all blocks within the same attention head and batch share $K$ and $V$ tensors. Co-locating these blocks within an XCD improves data reuse and reduces cross-device traffic.

\begin{figure}[h!]
    \centering
    \includegraphics[width=0.9\linewidth]{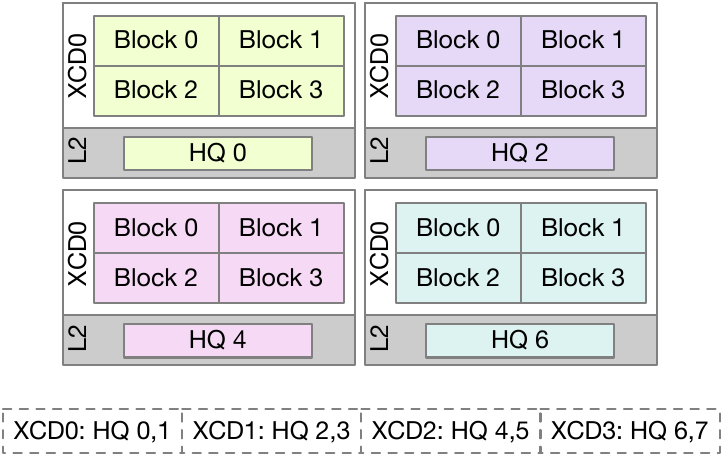}
    \caption{Swizzled Head-first mapping (eight qheads, 128 row blocks, four XCDs): Head-first iteration with spatial swizzling confines each attention head to a single XCD, maximizing cache locality and ensuring each XCD services one ACC at a time. Each unique color corresponds to a unique attention head.}
    \label{fig:swizzled-head-first}
\end{figure}

We introduce a novel mapping strategy, \textbf{Swizzled Head-first}, that uses spatial swizzling to assign all blocks of an attention head to the same XCD before moving to the next head. This approach uses spatial swizzling to ensure that workgroups within an ACC access shared data from one local cache, and XCDs service one ACC at a time. This strategy addresses the limitations of existing approaches by combining head-first iteration with spatial swizzling to co-locate all blocks of an attention head within a single XCD (Figure~\ref{fig:swizzled-head-first} and corresponding swizzling code in Figure~\ref{fig:head-first-code}).

\begin{figure}[h!]
    \centering
    \begin{lstlisting}[language=Python, 
                       basicstyle=\footnotesize\ttfamily,
                       breaklines=true,
                       backgroundcolor=\color{gray!10},
                       numbers=left,
                       numberstyle=\tiny,
                       numbersep=5pt,
                       showstringspaces=false,
                       commentstyle=\color{green!50!black},
                       keywordstyle=\color{blue},
                       stringstyle=\color{red}]
# Grid definition
grid = lambda META: (batch * num_q_heads * (seqlen_q/META["BLOCK_M"]))

# Swizzling 
wid = t1.program_id(0)  # workgroup id
wid_per_batch = wid // BATCH
heads_per_xcd = NUM_Q_HEADS // NUM_XCD
blocks_per_head = (SEQLEN_Q + BLOCK_M - 1) // BLOCK_M
chunk_size = NUM_XCD * blocks_per_head

# Calculate offsets
head_offset = ((wid_per_batch % NUM_XCD) * heads_per_xcd + wid_per_batch // (NUM_XCD * blocks_per_head))
block_offset = (wid_per_batch % chunk_size) // NUM_XCD
batch_offset = (wid // (blocks_per_head * NUM_Q_HEADS)) % BATCH
    \end{lstlisting}
    \caption{Workgroup ID swizzling logic for Swizzled Head-first Mapping. The swizzling scheme maps the linear workgroup ID (\texttt{wid}) to batch, head, and block offsets, ensuring attention heads from different ACCs are distributed across XCDs while sequence blocks within each ACC maintain locality on the same XCD.}
    \label{fig:head-first-code}
\end{figure}

Remarkably, this intuitive strategy requires minimal code modifications, as shown in Figure~\ref{fig:head-first-code}, yet proves highly effective at exploiting spatial locality. The head-first mapping consistently delivers improved L2 cache hit rates and substantial performance improvements, demonstrating the significant impact of NUMA-aware kernel design on multi-die chiplet-based GPUs.
\section{Evaluation}

\subsection{Experimental Setup}
\begin{table}[t]
\centering
\caption{AMD MI300X Architecture Specifications}
\label{tab:mi300x-specs}
\begin{tabular}{ll}
\toprule
\textbf{Component} & \textbf{Specification} \\
\midrule
\multicolumn{2}{l}{\textit{Compute Architecture}} \\
\quad Number of XCDs & 8 \\
\quad Compute Units per XCD & 38 (304 total) \\
\quad Stream Processors per CU & 64 \\
\midrule
\multicolumn{2}{l}{\textit{Memory Hierarchy}} \\
\quad L1 Cache per CU & 16 KB \\
\quad L2 Cache per XCD & 4 MB (32 MB total) \\
\quad HBM3 Capacity & 192 GB \\
\quad HBM3 Bandwidth & 5.3 TB/s \\
\bottomrule
\end{tabular}
\end{table}

We evaluate our spatial locality optimizations for FlashAttention2 forward pass on an AMD MI300X GPU, a chiplet-based architecture specifically designed for AI workloads. Table~\ref{tab:mi300x-specs} summarizes the key architectural characteristics relevant to our NUMA-aware optimizations. The MI300X features eight XCDs or Accelerator Complex Dies, with each XCD containing its own compute units, L2 cache, and memory controllers. This disaggregated memory hierarchy creates the NUMA effects that our optimization targets.


All kernels are implemented in Triton~\cite{triton}. We use ROCProfiler v3~\cite{rocprofv3} to measure L2 cache hit rates via hardware performance counters, specifically by monitoring the aggregated L2 cache hit rate across all XCDs.

\subsection{Evaluation Methodology}

\begin{figure*}[h]
    \centering
    \includegraphics[width=0.99\linewidth]{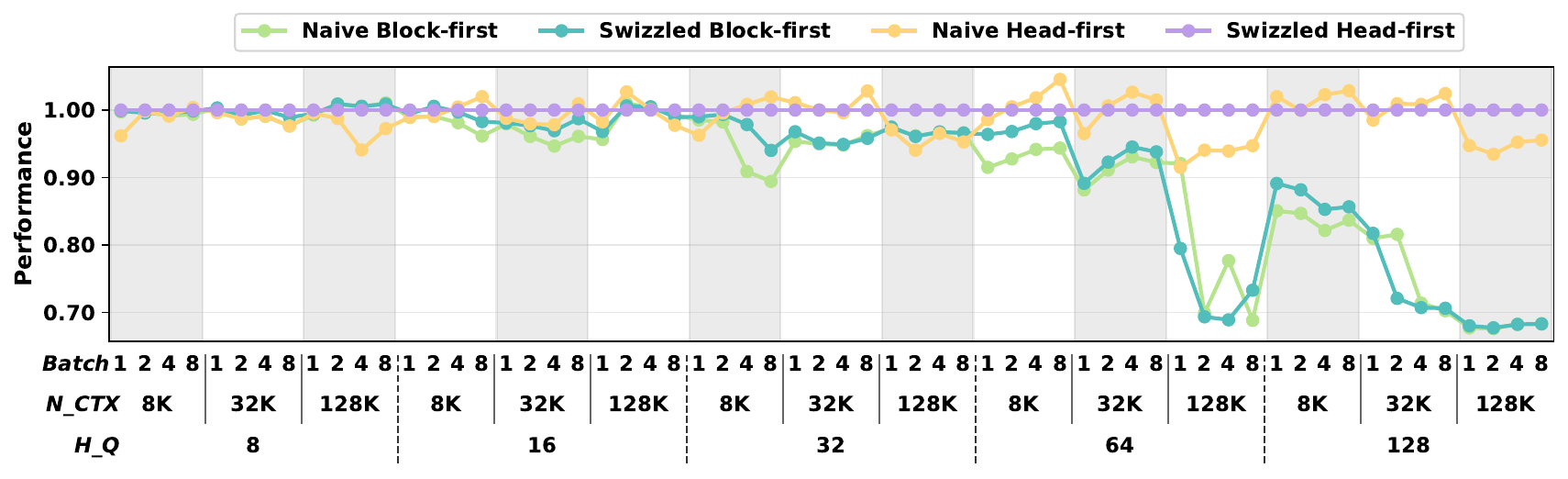}
    \vspace{-10pt}
    \caption{MHA Performance relative to Swizzled Head-first baseline across varying batch sizes (1-8) and sequence lengths (8K-128K). Block-first approaches show significant performance gaps that widen at higher head counts ($H\_Q \geq 64$), longer sequences, and larger batch sizes.}
    \label{fig:perfmha}
\end{figure*}

\begin{figure*}[h]
    \centering
    \includegraphics[width=0.99\linewidth]{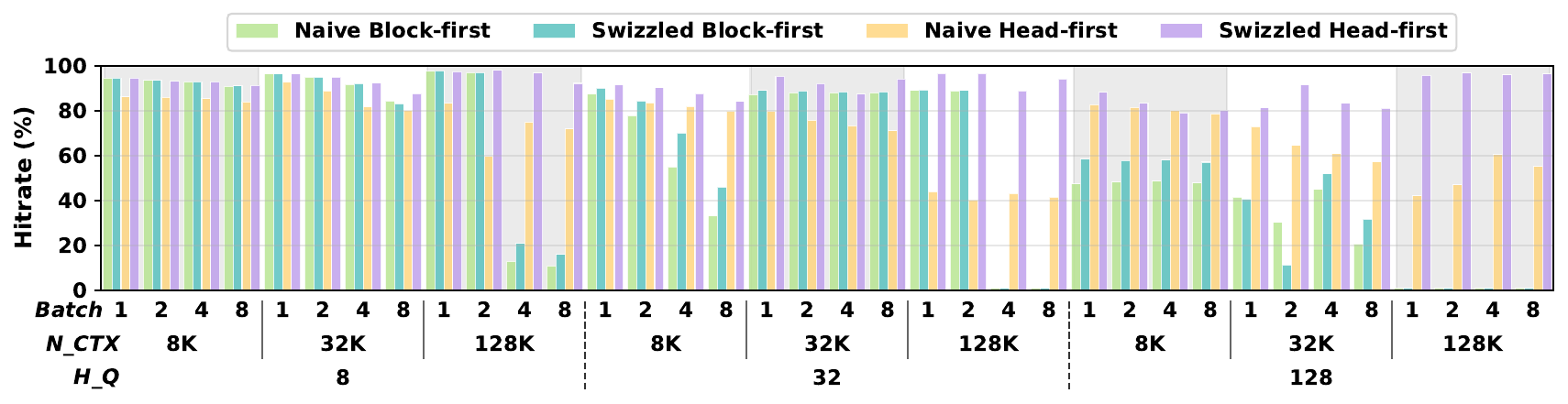}
    \vspace{-10pt}
    \caption{L2 Cache hit rates for MHA across varying batch sizes (1-8) and sequence lengths (2K-128K). The Swizzled Head-first approach consistently achieves high L2 hit rates across all configurations, whereas the other approaches exhibit drastic drops in hit rates at higher numbers of heads, longer sequence lengths, and higher batch sizes.}
    \label{fig:l2mha}
\end{figure*}

Our experimental methodology comprises three components: (1) a sensitivity analysis across MHA model hyperparameters to understand how sequence length, head count, and head dimension affect the benefits of our approach; (2) performance validation on GQA configurations to demonstrate generalization beyond MHA; and (3) a focused case study on the prefill stage of the DeepSeekV3 forward pass, representing a real-world deployment scenario with 128 attention heads.

We compare all four mapping strategies introduced in Section~\ref{sec:mapping}: Naive Block-first (Figure~\ref{fig:naive-block-first}), Swizzled Block-first (Figure~\ref{fig:swizzled-block-first}), Naive Head-first (Figure~\ref{fig:naive-head-first}), and our proposed Swizzled Head-first (Figure~\ref{fig:swizzled-head-first}).
For each experiment, performance metrics are normalized to the Swizzled Head-first approach, which represents the best-performing configuration across our experiments, enabling direct assessment of performance degradation in alternative approaches. Additionally, for the MHA sensitivity study, we measure L2 cache hit rates to validate the impact of improved cache locality on performance.

\begin{figure*}[h]
    \centering
    \includegraphics[width=0.99\linewidth]{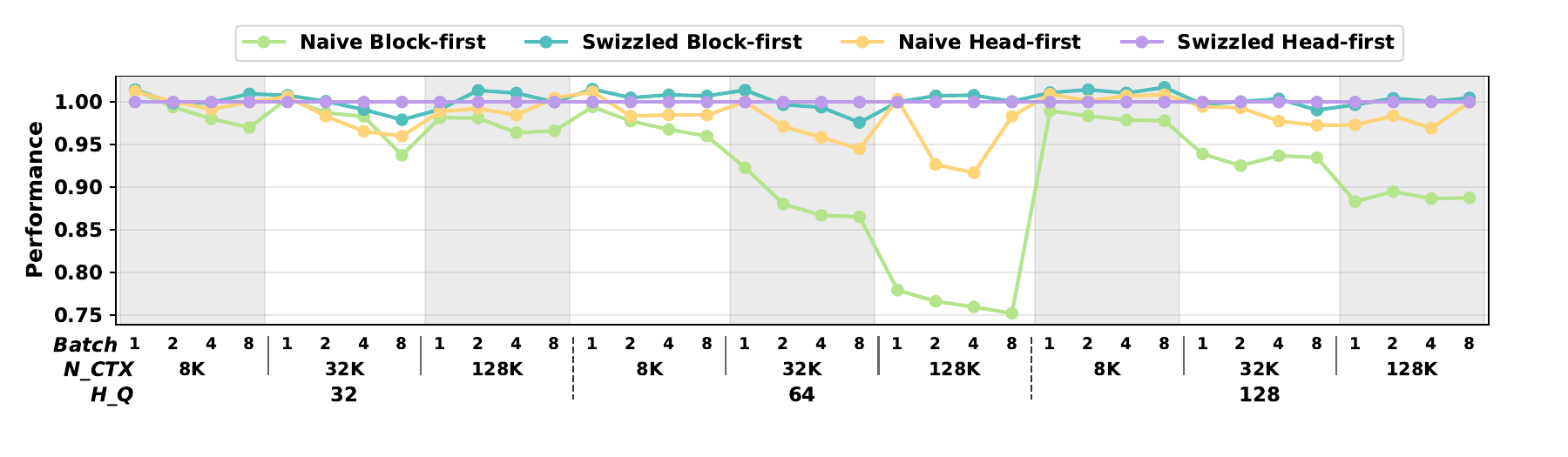}
    \vspace{-20pt} 
    \caption{Performance comparison for Grouped Query Attention with eight KV heads, normalized to Swizzled Head-first baseline. Results for $H\_Q = 32, 64, 128$ (corresponding to Llama 3 8B, 70B, 405B models) across sequence lengths 8K-128K and batch sizes 1-8. Both swizzled approaches achieve similar performance, while Naive Block-first mapping shows substantial degradation at higher attention head counts and longer sequences.}
    \label{fig:perfgqa}
\end{figure*}

\subsection{Multi-Head Attention Sensitivity Study}
We evaluated the four MHA mapping schemes across varying configurations to assess the impact of spatial locality on performance and L2 cache utilization. Our experiments sweep across number of attention heads~(8 to 128), sequence lengths~(2K to 128K tokens), batch sizes~(1 to 8), and used fixed BLOCK\_M = 128 and BLOCK\_N = 64 (Table~\ref{tab:mha_sweep}).

\begin{table}[t]
\centering
\caption{Experimental configuration for Multi-Head Attention (MHA) Sensitivity Study}
\label{tab:mha_sweep}
\begin{tabular}{lc}
\toprule
\textbf{Parameter} & \textbf{Values} \\
\midrule
Context Length ($N\_{CTX}$) & 8K, 32K, 128K \\
Batch Size & 1, 2, 4, 8 \\
Number of Heads ($H\_Q = H\_K$) & 8, 16, 32, 64, 128 \\
Head Dimension ($D\_{HEAD}$) & 128 \\
Block Size ($M \times N$) & $128 \times 64$ \\
\bottomrule
\end{tabular}
\end{table}

Figure~\ref{fig:perfmha} demonstrates that the Swizzled Head-first approach, our proposed method, serves as the performance baseline. For a smaller number of heads, all approaches perform similarly. The block-first approaches show significant performance degradation across most configurations, with the gap widening as the number of heads and sequence length increase. The performance disadvantage is most pronounced at $H_Q \geq 64$ with longer sequences ($N_{CTX} \geq 32K$), where Block-first approaches achieve only 64-70\% the efficiency of our Swizzled Head-first method. At $H_Q = 128$ with $N_{CTX} = 128K$, the Swizzled Head-first approach achieves up to 50\% higher performance than Block-first mappings. Interestingly, the Naive Head-first mapping's performance closely matches our swizzled approach for most cases, except at very high sequence lengths, where it drops to $\sim90\%$ relative efficiency. This observation aligns with our theory that the Naive Head-first approach still benefits from cache locality by having the relevant ACC's data in the caches, even if it's replicated across all caches, except at very high sequence lengths, where cache splitting with different ACC's data and redundant memory fetches degrade performance.

The L2 cache hit rate analysis in Figure~\ref{fig:l2mha} reveals the underlying cause of these performance differences. With fewer heads and shorter sequences, all approaches maintain high L2 hit rates ($\sim$90\%). However, cache behavior diverges dramatically as workload complexity increases. At H\_Q = 128 with N\_CTX = 128K, the Swizzled Head-first mapping sustains 90-96\% hit rates across all batch sizes, the hit-rate of Naive Head-first falls to 40-60\%, while block-first approaches collapse to approximately 1\% hit rates. Even the Swizzled Block-first method, which is deployed in current GPU kernels, almost always misses in cache at these extreme configurations, despite performing better at moderate scales. This catastrophic cache degradation directly explains the 45-50\% performance gap observed in Figure~\ref{fig:perfmha}.

\begin{figure*}[h]
    \centering
    \includegraphics[width=0.99\linewidth]{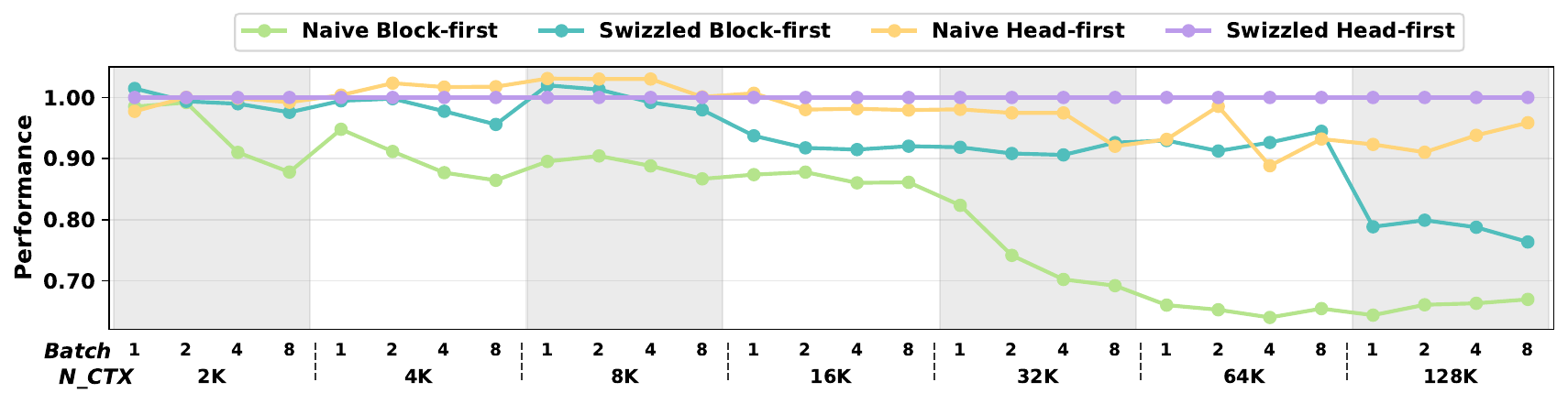}
    \vspace{-10pt}
    \caption{DeepSeekV3 prefill performance with 128 attention heads, normalized to Swizzled Head-first baseline, across sequence lengths 2K-128K and batch sizes 1-8. Block-first approaches show significant performance degradation at longer sequences, with Naive Block-first dropping below $0.65\times$ relative efficiency at 128K tokens.}
    \label{fig:dsmhaperf}
\end{figure*}

\begin{table}[t]
\centering
\caption{Model configurations for Llama Models (GQA) and DeepSeek-V3 (MHA)}
\label{tab:model_params}
\begin{tabular}{lcccc}
\toprule
\textbf{Model} & \textbf{Attn. Type} & \textbf{H\_Q} & \textbf{H\_K} & \textbf{D\_HEAD}\\
\midrule
Llama-3 8B & GQA & 32 & 8 & 128  \\
Llama-3 70B & GQA & 64 & 8 & 128 \\
Llama-3 405B & GQA & 128 & 8 & 128 \\
\midrule
DeepSeek-v3 & MHA & 128 & 128 & 56 \\
\bottomrule
\end{tabular}
\end{table}

This substantial difference in cache utilization directly translates to the observed performance benefits, as reduced cache misses minimize expensive off-chip memory accesses. The Swizzled Head-first strategy's ability to maintain high cache utilization under extreme conditions directly translates to sustained computational throughput, making it essential for scaling attention mechanisms to large head counts and long context lengths.

\subsection{Grouped Query Attention Sensitivity Study}

\begin{figure}[h]
    \centering
    \includegraphics[width=\linewidth]{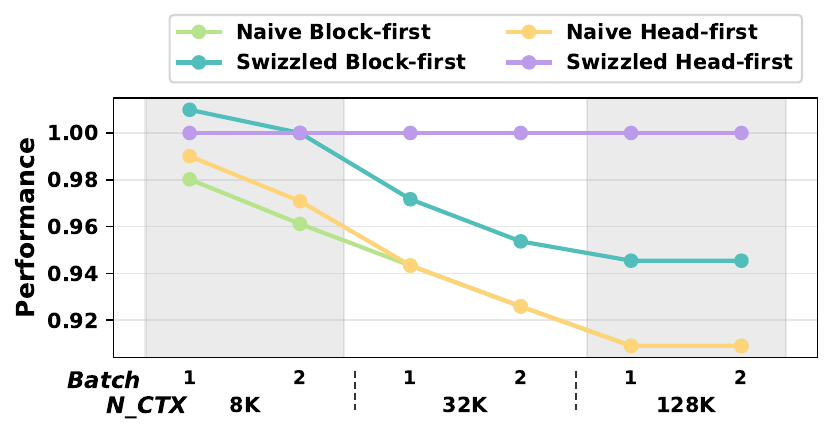}
    \vspace{-10pt}
    \caption{Backward pass speedup comparison of mapping schemes with 128 query heads versus the Naive Block-first approach. Results shown across context lengths from 8K to 128K tokens and batch sizes 1 to 8. The Swizzled Head-first approach demonstrates increasing speedup with longer sequences, achieving 1.10$\times$ at 128K tokens.}
    \label{fig:bwperf}
    
\end{figure}

We evaluated the four mapping schemes on GQA configurations with a fixed eight KV heads and varying query heads (H\_Q = 32, 64, 128), corresponding to the Llama 3 model family sizes (8B, 70B, and 405B parameters, respectively). As multiple query heads share the same KV heads, GQA creates different memory access patterns and cache pressure characteristics compared to MHA.

Figure~\ref{fig:perfgqa} demonstrates that the Swizzled Head-first and Swizzled Block-first approaches maintain robust performance across all GQA configurations. At H\_Q = 32, all mapping schemes achieve similar performance. The Naive Block-first method exhibits significant degradation at higher query heads, sequence lengths, and batch sizes, since the L2 cache in each XCD is split across different ACCs' data.

The Swizzled Block-first approach performs well with GQA, as the eight KV heads match the number of XCDs in the MI300X architecture, avoiding any cache splits. In the case of Swizzled Head-first, each ACC, which is now a group of all query heads sharing the KV data, is mapped to one XCD, achieving similarly strong performance. The Naive Head-first approach shows instability at higher sequence lengths and batches, with performance dipping to about 90-95\% of the Swizzled Head-first approach in some cases, indicating non-ideal cache utilization and higher redundant memory accesses. These results confirm that spatial locality optimizations remain effective for GQA workloads.

\subsection{Case Study: Optimizing DeepSeekV3 Prefill}
DeepSeekV3's prefill phase uses MHA with 128 query heads and 128 key-value heads, making it an ideal candidate for evaluating our Swizzled Head-first optimization, as the number of attention heads (128) significantly exceeds the number of XCDs (8) available on the MI300X GPU.

As shown in Figure~\ref{fig:dsmhaperf}, the Swizzled Head-first approach achieves superior performance across most configurations, particularly at longer sequence lengths. At extreme sequence lengths, the Naive Block-first method shows the worst performance, degrading to under 65\% relative performance at 128K tokens. The Swizzled Block-first method also degrades substantially to 76\% of the Swizzled Head-first method at 128K tokens with a batch size of eight. The Naive Head-first approach also shows instability at extreme lengths, with performance varying significantly. 
The smaller head dimension in this configuration (D\_HEAD=56) reduces overall arithmetic intensity, thereby lowering absolute performance across all methods.

These results confirm that the Swizzled Head-first optimization provides substantial and scalable performance benefits for DeepSeekV3's high-head-count MHA configuration.


\subsection{FlashAttention2 Backward Pass}
We evaluated the four mapping schemes for the FlashAttention2 backward pass kernel in AMD's AITER~\cite{aiter} repository with H\_Q = 128 attention heads across varying context lengths (8K, 32K, 128K) and batch sizes (1, 2). Figure~\ref{fig:bwperf} shows that the Swizzled Head-first approach consistently outperforms other mapping schemes, with the performance of the Swizzled Block-first approach degrading to about $0.94 \times$ and the performance of the Naive Block-first and Naive Head-first approaches degrading to $0.91\times$ at 128K tokens. Given the additional complexity and scalar operations required in the FlashAttention2 backward pass, we suspect that further gains may be constrained by emerging bottlenecks introduced by the Swizzled Head-first optimization. This behavior aligns with our observations across other kernels, where performance improvements of up to 50\% were achieved. We leave further optimization to future work.

\section{Conclusion}
This work demonstrates that spatial locality in attention workloads can be strategically exploited to mitigate NUMA effects in chiplet-based GPUs with disaggregated memory. We identify inherent spatial locality patterns in the attention mechanism and introduce the Swizzled Head-first Mapping strategy, which aligns computation with the NUMA domains of multi-die GPU architectures. On AMD's MI300X architecture, our approach achieves up to 50\% performance improvement over state-of-the-art attention algorithms using conventional scheduling techniques in multi-head attention while maintaining consistently high L2 cache hit rates of 80-97\%. These results, achieved with minimal code changes, demonstrate that significant performance gains can be obtained via architecture-aware kernel design. Our findings underscore the necessity of hardware-aware algorithm design as chiplet-based architectures become increasingly prevalent in AI accelerators.

\bibliography{ref}
\bibliographystyle{mlsys2025}

%


\end{document}